\input{aipcheck}

\documentclass[
    ,final            
  ]
  {aipproc}

\layoutstyle{6x9}

\usepackage{amsfonts,amsmath,amssymb,amsthm,latexsym}
\usepackage[all]{xy}

\theoremstyle{definition}

\theoremstyle{remark}

\begin{document}

\title[Super-Lie algebra realizations via paraparticles]{Super-Hopf realizations of Lie superalgebras: Braided Paraparticle extensions of the Jordan-Schwinger map}


\author{K. Kanakoglou}{
  address={Instituto de F\'{\i}sica y Matem\'{a}ticas (\textsc{Ifm}),
Universidad Michoacana de San Nicolas de Hidalgo (\textsc{Umsnh}),
Edificio C-3, Cd. Universitaria, CP 58040,
Morelia, Michoac\'{a}n, \textsc{Mexico}}, altaddress={School of Physics, Nuclear and Elementary Particle Physics Department, Aristotle University of Thessaloniki (\textsc{Auth}),
54124 Thessaloniki, \textsc{Greece}}
}

\author{C. Daskaloyannis}{
  address={School of Mathematics, Analysis Department, Aristotle University of Thessaloniki (\textsc{Auth}), 54124 Thessaloniki, \textsc{Greece}}
}

\author{A. Herrera-Aguilar}{
  address={Instituto de F\'{\i}sica y Matem\'{a}ticas (\textsc{Ifm}),
Universidad Michoacana de San Nicolas de Hidalgo (\textsc{Umsnh}),
Edificio C-3, Cd. Universitaria, CP 58040,
Morelia, Michoac\'{a}n, \textsc{Mexico}}
}

\begin{abstract}
The mathematical structure of a mixed paraparticle system (combining both parabosonic and parafermionic degrees of
freedom) commonly known as the Relative Parabose Set, will be investigated and a braided group structure will be
described for it. A new family of realizations of an arbitrary Lie superalgebra will be presented and it will be
shown that these realizations possess the valuable representation-theoretic property of transferring invariably the
super-Hopf structure. Finally two classes of virtual applications will be outlined: The first is of interest for both
mathematics and mathematical physics and deals with the representation theory of infinite dimensional Lie superalgebras,
while the second is of interest in theoretical physics and has to do with attempts to determine specific classes of
solutions of the Skyrme model.
\end{abstract}

\maketitle {\small \textbf{\underline{Published at:}} \textsc{Aip} Conf. Proc. v. \textbf{1256}, p.193-200, 2010}


\section{Introduction}

{\small
The idea of realizing an algebra of operators in terms of an other algebra,
roughly amounts in constructing a suitable homomorphism, i.e. a linear map preserving the multiplicative relations, between our initial algebra and the ``target'' algebra. In this way, the relations of the initial algebra can be directly reproduced by the relations of the target algebra (which are either easier to handle or more well known). Moreover, the representations
of the target algebra give rise -in a straightforward way- to representations of the initial algebra, enabling us to gain information about the spectrum, the eigenstates and thus the expected values of various physical quantities. The dynamical variables of any physical theory frequently have the structure of the universal enveloping algebra (\textsc{Uea}) of some Lie algebra and this is exactly what initiated the interest in realizations of Lie algebras. Historically, one of the earliest applications of such a methodology was the famous Jordan-Schwinger map \cite{Jor} which was first introduced in $1935$ by P. Jordan:
$$
\begin{array}{llll}
x \stackrel{J_{B}}{\mapsto} J_{B}(x) \equiv \sum_{i,j=1}^{n} E_{ij}(x)b_{i}^{+}b_{j}^{-}  & & & x \stackrel{J_{F}}{\mapsto} J_{F}(x) \equiv \sum_{i,j=1}^{n} E_{ij}(x)f_{i}^{+}f_{j}^{-}
\end{array}
$$
$J_{B}: U(L) \rightarrow B$ and $J_{F}: U(L) \rightarrow F$ are (associative) algebra homomorphisms from the (initial) \textsc{Uea} $U(L)$ of $L$ to the
(target) bosonic (CCR) $B$ or fermionic (CAR) $F$ algebras respectively. $L$ is any Lie algebra with a fin. dim. $n^{2}$-matrix representation $E$, $U(L)$ is the \textsc{Uea} of $L$ and $x$ is any element of $L$ (thus $x$ is a generator of $U(L)$). As is well known, the boson-fermion versions $J_{B}, J_{F}$ of the Jordan-Schwinger map, led to the construction of the symmetric and the antisymmetric representations respectively of the Lie algebras which were followed by extended applications in various areas of physics. For enlightening reviews regarding the history and the usefulness of the Jordan-Schwinger realization in physics problems see \cite{BiLo}.
The success and the applicability of the J.S. map inspired the extension of the idea and the construction of other realizations as well using bosonic or fermionic degrees of freedom as target algebras. The relevant examples, of realizing the dynamical variables of some model via bosonic or fermionic degrees of freedom, are numerous: In the review article \cite{KlMa} and the references therein one can find a host of boson-fermion realizations and their applications in models of nuclear physics, while in \cite{Ca} such techniques are applied to models from  various areas of physics.

The introduction of paraparticle algebras \cite{GreeMe} and the study of their representations in connection to the usual boson-fermion Fock representations, gave the occasion for the study of \emph{realizations of Lie algebras with parabosons or parafermions} \cite{DaKa, Kad} which may be considered as formal (paraparticle) generalizations of the J.S. map.

Since the introduction of the idea of supersymmetry, Lie superalgebras and their \textsc{Uea}s have dramatically come into play. Nowadays the Lie superalgebraic structure can be found throughout models of quantum field theory, elementary particle physics, solid state physics etc. Although Lie superalgebra realizations is a much more recent topic than Lie algebra realizations, there is already a significant number of references on this subject. Far from trying to present an exhaustive list, we feel it's worth underlining the readers attention to the general works \cite{Hong} -some of which may be considered as formal (superalgebraic) generalizations of the J.S. map- and the references therein. In these works, and to the best of our knowledge in any similar reference in the literature, the target algebra is almost always some mixture of bosons and fermions (ordinary or deformed). In some of these references, algebras of differential operators -acting on suitable polynomial spaces- have also been used. We can thus speak about \emph{particle realizations} or \emph{differential realizations} of Lie superalgebras.

The novel thing in this article will be the use of a mixed paraparticle system as a target algebra. Consequently we will deal with \emph{paraparticle realizations} of an arbitrary Lie superalgebra, a topic which is almost totally unexplored in the literature. More specifically, our target algebra will be a system combining both parabosonic and parafermionic degrees of freedom be the so-called \emph{Relative Parabose Set} $P_{BF}$, according to the terminology introduced by Greenberg in \cite{GreeMe}. Let us close this introduction by presenting the multiplicative structure of $P_{BF}$.
\paragraph{\textbf{Multiplicative structure of the Relative Parabose Set $P_{BF}$}}
The Relative Parabose Set has been historically the only -together with the \emph{Relative Parafermi Set} $P_{FB}$- attempt for a mixture of interacting parabosonic and parafermionic degrees of freedom. We present it here in terms of generators and relations, adopting a handy notation: $P_{BF}$ is generated -as an assoc. alg.- by the (infinite) generators $B_{i}^{\xi}$, $F_{j}^{\eta}$, for all values $i,j = 1, 2, ...$ and $\xi, \eta = \pm$. The relations satisfied by the above generators are:

The usual trilinear relations of the parabosonic and the parafermionic algebras which can be compactly summarized as
\small{\begin{equation} \label{parab-paraf}
\begin{array}{c}
\big[ \{ B_{i}^{\xi},  B_{j}^{\eta}\}, B_{k}^{\epsilon}  \big] = (\epsilon - \eta)\delta_{jk}B_{i}^{\xi} + (\epsilon -
 \xi)\delta_{ik}B_{j}^{\eta}    \\
\big[ [ F_{i}^{\xi},  F_{j}^{\eta} ], F_{k}^{\epsilon}  \big] = \frac{1}{2}(\epsilon - \eta)^{2} \delta_{jk}F_{i}^{\xi} - \frac{1}{2}(\epsilon -
 \xi)^{2} \delta_{ik}F_{j}^{\eta}   \\
 \end{array}
\end{equation}}
for all values $i, j, k = 1, 2, ..., $ and $\xi, \eta, \epsilon = \pm$, \ together with the mixed trilinear relations
\begin{equation} \label{parabparaf}
\begin{array}{c}
\big[ \{ B_{k}^{\xi},  B_{l}^{\eta}\}, F_{m}^{\epsilon}  \big] = \big[ [ F_{k}^{\xi},  F_{l}^{\eta} ], B_{m}^{\epsilon}  \big] = 0    \\
\big[ \{ F_{k}^{\xi},  B_{l}^{\eta}\}, B_{m}^{\epsilon}  \big] = (\epsilon - \eta) \delta_{lm} F_{k}^{\xi},  \ \ \ \
\big\{ \{ B_{k}^{\xi},  F_{l}^{\eta}\}, F_{m}^{\epsilon}  \big\} = \frac{1}{2}(\epsilon - \eta)^{2} \delta_{lm} B_{k}^{\xi}
\end{array}
\end{equation}
for all values $k, l, m = 1, 2, ..., $ and $\xi, \eta, \epsilon = \pm$, \ which represent a kind of algebraically established interaction between parabosonic and parafermionic degrees of freedom and characterize the relative parabose set.

One can easily observe that the relations \eqref{parab-paraf} involve only the parabosonic and the parafermionic degrees of freedom separately while the ``interaction'' relations \eqref{parabparaf} mix the parabosonic with the parafermionic degrees of freedom according to the recipe proposed in \cite{GreeMe}. In all the above and in what follows, we use the notation $[x, y]$ (i.e.: the ``commutator'') to imply the expression $xy-yx$ and the notation $\{x, y \}$ (i.e.: the ``anticommutator'') to imply the expression $xy+yx$, for $x$ and $y$ any elements of the algebra $P_{BF}$.

Finally we remark that all vector spaces, algebras and tensor products in this article will be considered over the field of complex numbers $\mathbb{C}$ and that the prefix ``super'' will always amount to $\mathbb{Z}_{2}$-graded.

\section{$(\mathbb{G}, \theta)$-Lie algebras: Colors and Braidings}

In this section $\mathbb{G}$ will always be considered a finite, abelian group.

A \emph{$(\mathbb{G}, \theta)$-Lie algebra} or a \emph{$\theta$-colored, $\mathbb{G}$-graded Lie algebra} \cite{Scheu2} is a mathematical idea generalizing the idea of Lie algebras and Lie superalgebras. It consists of a $\mathbb{G}$-graded vector space $L$ i.e. $L = \oplus_{g \in \mathbb{G}} L_{g}$ (equiv.: a representation of the group Hopf algebra $\mathbb{CG}$ on the vector space $L$), a non-associative multiplication $\langle .., .. \rangle : L \times L \rightarrow L$ on $L$ respecting the gradation, in the sense that $\langle L_{a}, L_{b} \rangle \subseteq L_{a+b}, \ \forall \ a,b \in \mathbb{G}$ and a function $\theta: \mathbb{G} \times \mathbb{G} \rightarrow \mathbb{C}^{*}$ taking non-zero complex values. The above data must be subject to the following set of axioms:
\begin{itemize}
\item $\theta$-\textbf{braided} ($\mathbf{\mathbb{G}}$-graded) antisymmetry: $\langle x, y \rangle = - \theta(a,b) \langle y, x \rangle$
\item $\theta$-\textbf{braided} ($\mathbf{\mathbb{G}}$-graded) Jacobi id.:
$
\theta(c,a) \langle x, \langle y, z \rangle \rangle + \theta(b,c) \langle z, \langle x, y \rangle \rangle +   \theta(a,b) \langle y, \langle z, x \rangle \rangle = 0
$
\item $\theta:\mathbb{G} \times \mathbb{G} \rightarrow \mathbb{C}^{*} \rightsquigarrow$
\textbf{color function} $\leftrightsquigarrow$
              $
              \left\{
                \begin{array}{c}
                  \theta(a+b, c) = \theta(a,c) \theta(b,c),  \\
                  \theta(a, b+c) = \theta(a,b) \theta(a,c),  \\
                  \theta(a,b) \theta(b,a) = 1
                \end{array}
              \right.
              $
\end{itemize}
for all homogeneous $x \in L_{a}$, $y \in L_{b}$, $z \in L_{c}$ and $\forall$ $a, b, c \in \mathbb{G}$.

The $\theta$ function is called a \emph{color function} on $\mathbb{G}$ or a \emph{commutation factor} on $\mathbb{G}$. The axioms imposed on it imply that it is a skew-symmetric \emph{bicharacter} on $\mathbb{G}$ which has been shown to be equivalent \cite{Mon} to a \emph{triangular universal $R$-matrix} on the group Hopf algebra $\mathbb{C}\mathbb{G}$. We thus conclude that by definition, the notion of $\theta$-colored $\mathbb{G}$-graded Lie algebra implicitly presupposes the existence of a triangular structure for the group Hopf algebra $\mathbb{C}\mathbb{G}$ which finally entails a \emph{symmetric braiding} in the monoidal category ${}_{\mathbb{CG}}\mathcal{M}$ of its representations. This last remark, fully justifies the use of the term \emph{braided}, in both the antisymmetry property and the generalized Jacobi identity included in the defining axioms of the $\theta$-colored $\mathbb{G}$-graded Lie algebra.

The \textsc{Uea} of $L$, $\mathbb{U}(L)$ is a $\mathbb{G}$-gr. assoc. algebra. It will be generated by any linearly independent set of homogeneous elements of $L$. Inside $\mathbb{U}(L)$ we will have the relations $\langle x, y \rangle = xy - \theta \big( deg(x),deg(y) \big) yx$, for any $x,y$ homogeneous in  $L$ and  $deg(x), deg(y) \in  \mathbb{G}$.

The $\mathbb{G}$-graded v.s. $\mathbb{U}(L) \otimes \mathbb{U}(L)$ equipped with the multiplication
$
(x \otimes y)(z \otimes w) = \theta(a,b) xz \otimes yw
$
is denoted $\mathbb{U}(L) \underline{\otimes} \mathbb{U}(L)$ and called \emph{$\mathbf{\theta}$-braided, $\mathbb{G}$-graded tensor product algebra}.

$\mathbb{U}(L)$ can be equipped with a ``comultiplication'' $\underline{\Delta} : \mathbb{U}(L) \rightarrow \mathbb{U}(L) \underline{\otimes} \mathbb{U}(L)$
which is an homomorphism of $\mathbb{G}$-gr. assoc. alg., uniquely defined by its values on the homog. elements of $L$
$
\underline{\Delta}(x) = 1 \otimes x + x \otimes 1
$
and with an ``antipode'' $\underline{S} : U(L) \rightarrow U(L)$ which is a \emph{twisted} \cite{Mon} or \emph{braided antihomomorphism} in the sense that:
$\underline{S}(xy) = \theta \big( deg(x), deg(y) \big) \underline{S}(y)\underline{S}(x)$ for any homogeneous $x, y \in U(L)$. $\underline{S}$ is uniquely defined by its values on the homogeneous elements of $L$: $\underline{S}(x) = -x$. If the above are supplemented with $\underline{\varepsilon}(x) = 0 $ $\ \forall$ $x \in \mathbb{U}(L)$ $\ x \neq 1$, $\underline{\varepsilon}(1) = 1$ then  all usual axioms of the Hopf structure (coassociativity, counity, antipode compatibility, etc) are satisfied.

The resulting algebraic structure is strongly reminiscent of a Hopf algebra but it is not one -at least not in the ordinary sense. Such structures are traditionally called in the literature as $\mathbb{G}$-graded Hopf algebras or $(\mathbb{G}, \theta)$-graded Hopf algebras. According to the modern terminology \cite{Mon}, developed in the '$90$'s and originating from Quantum Group theory, such a structure will  be called a $\theta$-\textbf{braided group} in the sense of the braiding induced in the ${}_{\mathbb{CG}}\mathcal{M}$ category by the commutation factor $\theta$. It is also customary to speak of $\mathbb{U}(L)$ as \emph{a Hopf algebra in the symmetric monoidal category ${}_{\mathbb{CG}}\mathcal{M}$ of representations of the group Hopf algebra $\mathbb{CG}$}. Lets proceed to a couple of examples.

\paragraph{\textbf{Lie superalgebras:}}
If we set $\mathbb{G} = \mathbb{Z}_{2}$ and the commutation factor
$$
\begin{array}{ccccc}
\theta: \mathbb{Z}_{2} \times \mathbb{Z}_{2} \rightarrow \mathbb{C}^{*}  & & given \ by & & \theta(a,b) = (-1)^{a \cdot b}
\end{array}
$$
($\forall$ $a, b \in \mathbb{Z}_{2}$, and the oper. in the exponent in the $\mathbb{Z}_{2}$ ring), we get a $\mathbb{Z}_{2}$-graded Lie algebra or a Lie superalgebra.
\begin{small}
Thus: $L = L_{0} \oplus L_{1}$, and in $\mathbb{U}(L)$, we have: $\mathbf{\langle x, y \rangle = xy - (-1)^{deg(x)deg(y)}yx}$ \\
($\forall \ x,y$ hom. in $L$) which imply:
$$
\begin{array}{c}
L_{0} \ni \langle L_{0}, L_{0} \rangle =  [ L_{0}, L_{0} ]  \rightsquigarrow  \textrm{commutator} \\
L_{1} \ni \langle L_{0}, L_{1} \rangle =  [ L_{0}, L_{1} ]  \rightsquigarrow  \textrm{commutator}       \\
L_{0} \ni \langle L_{1}, L_{1} \rangle =  \{ L_{1}, L_{1} \}   \rightsquigarrow \textrm{anticommutator}      \\
\end{array}
$$
\end{small}

\paragraph{\textbf{$\mathbf{(\mathbb{Z}_{2} \times \mathbb{Z}_{2})}$-graded Lie algebras:}}
If we set $\mathbb{G} = \mathbb{Z}_{2} \times \mathbb{Z}_{2}$ and the commutation factor
$$
\begin{array}{ccccc}
\theta:\big( \mathbb{Z}_{2} \times \mathbb{Z}_{2} \big) \times \big( \mathbb{Z}_{2} \times \mathbb{Z}_{2} \big) \rightarrow \mathbb{C}^{*}
& & given \ by & & \theta(a,b) = (-1)^{(a_{1} \cdot b_{1} + a_{2} \cdot b_{2})}
\end{array}
$$
($\forall$ $a = (a_{1}, a_{2}), b = (b_{1}, b_{2}) \in \mathbb{Z}_{2} \times \mathbb{Z}_{2}$, and the oper. in the exponent are  in the
$\mathbb{Z}_{2}$ ring), we get (one of the possible) examples of a $(\mathbb{Z}_{2} \times \mathbb{Z}_{2})$-graded Lie algebra.
\begin{small}
Thus: $L = L_{00} \oplus L_{01} \oplus L_{10} \oplus L_{11}$, and
in $\mathbb{U}(L)$, we have: $\mathbf{\langle x, y \rangle = xy -} \theta(deg(x),deg(y))\mathbf{yx}$ ( $\forall \ x,y$ hom. in $L$) which imply:
$$
\begin{array}{cc}
 L_{00} \ni \langle L_{00}, L_{00} \rangle =  [ . , . ]  \rightsquigarrow  commutator & L_{10} \ni \langle L_{00}, L_{10} \rangle =  [ . , . ]  \rightsquigarrow commutator  \\
 L_{00} \ni \langle L_{01}, L_{01} \rangle =  \{ . , . \}  \rightsquigarrow anticommutator  &  L_{11} \ni \langle L_{00}, L_{11} \rangle =  [ . , . ]  \rightsquigarrow commutator \\
 L_{00} \ni \langle L_{10}, L_{10} \rangle =  \{ . , . \}  \rightsquigarrow anticommutator &   L_{11} \ni \langle L_{01}, L_{10} \rangle =  [ . , . ]  \rightsquigarrow commutator \\
 L_{00} \ni \langle L_{11}, L_{11} \rangle =  [ . , . ]  \rightsquigarrow commutator &   L_{10} \ni \langle L_{01}, L_{11} \rangle =  \{ . , . \}  \rightsquigarrow anticommutator \\
 L_{01} \ni \langle L_{00}, L_{01} \rangle =  [ . , . ]  \rightsquigarrow commutator &   L_{01} \ni \langle L_{10}, L_{11} \rangle =  \{ . , . \}  \rightsquigarrow anticommutator
\end{array}
$$
\end{small}

\section{The Relative Parabose Set as a braided group}

In this section we review results presented in \cite{KaDaHa, Ya1} regarding the braided, graded algebraic structure of the Relative Parabose Set $P_{BF}$.

\paragraph{\textbf{$\mathbf{(\mathbb{Z}_{2} \times \mathbb{Z}_{2})}$-Graded structure of $P_{BF}$}}

The relative parabose set $P_{BF}$ is the universal enveloping algebra UEA of a
$\theta$-colored ($\mathbb{Z}_{2} \times \mathbb{Z}_{2}$)-graded Lie algebra $L_{\mathbb{Z}_{2} \times \mathbb{Z}_{2}}$.
This implies that $P_{BF}$ is a ($\mathbb{Z}_{2} \times \mathbb{Z}_{2}$)-graded associative algebra
\begin{equation}
P_{BF} \cong \mathbb{U}(L_{\mathbb{Z}_{2} \times \mathbb{Z}_{2}})
\end{equation}
Its generators are homogeneous elements in the above gradation, with the paraboson generators $B_{k}^{+}$, $B_{l}^{-}$,
$k,l = 1, 2, ...$ spanning the $L_{10}$ subspace of $L_{\mathbb{Z}_{2} \times \mathbb{Z}_{2}}$, and the parafermion generators
$F_{\alpha}^{+}$, $F_{\beta}^{-}$, $\alpha, \beta = 1, 2, ...$ spanning the $L_{11}$ subspace of $L_{\mathbb{Z}_{2} \times \mathbb{Z}_{2}}$, thus
their grades are given as follows
\begin{equation} \label{gradPBF1}
\begin{array}{ccc}
deg(B_{k}^{\varepsilon}) = (1,0) & & deg(F_{\alpha}^{\eta}) = (1,1)
\end{array}
\end{equation}
where $\varepsilon, \eta = \pm$. At the same time the polynomials $\{ B_{k}^{\epsilon}, B_{l}^{\eta} \}$ and $[ F_{\alpha}^{\epsilon}, F_{\beta}^{\eta} ]$ $\forall$ $k, l, \alpha, \beta = 1,2,... \ $ and $\forall$ $\epsilon, \eta = \pm$ span the subspace $L_{00}$ of $L_{\mathbb{Z}_{2} \times \mathbb{Z}_{2}}$,
and the polynomials $\{ F_{\alpha}^{\epsilon}, B_{k}^{\eta} \}$ $\forall$ $k, \alpha = 1,2,... \ $ and $\forall$ $\epsilon, \eta = \pm$ span the subspace $L_{01}$ of $L_{\mathbb{Z}_{2} \times \mathbb{Z}_{2}}$. Consequently their grades are given as follows
\begin{equation}  \label{gradPBF2}
\begin{array}{ccc}
deg(\{ B_{k}^{\epsilon}, B_{l}^{\eta} \}) = deg([ F_{\alpha}^{\epsilon}, F_{\beta}^{\eta} ]) = (0,0) &  &
deg(\{ F_{\alpha}^{\epsilon}, B_{k}^{\eta} \}) = (0,1)
\end{array}
\end{equation}
Finally the color function used in the above construction is given
\begin{equation} \label{colfunctRelParSe}
\begin{array}{c}
\theta:\big( \mathbb{Z}_{2} \times \mathbb{Z}_{2} \big) \times \big( \mathbb{Z}_{2} \times \mathbb{Z}_{2} \big) \rightarrow \mathbb{C}^{*} \\
  \\
\theta(a,b) = (-1)^{(a_{1}b_{1} + a_{2}b_{2})}
\end{array}
\end{equation}
for all $a = (a_{1}, a_{2}), b = (b_{1}, b_{2}) \in \mathbb{Z}_{2} \times \mathbb{Z}_{2}$, and the operations in the exponent are considered in the
$\mathbb{Z}_{2}$ ring.

\paragraph{\textbf{Braided Group structure of $P_{BF}$}}

The relative parabose set $P_{BF}$ has the structure of a $\theta$-braided group where the commutation factor $\theta:\big( \mathbb{Z}_{2} \times \mathbb{Z}_{2} \big)
\times \big( \mathbb{Z}_{2} \times \mathbb{Z}_{2} \big) \rightarrow \mathbb{C}^{*}$ is given by \eqref{colfunctRelParSe}. The relations can be given explicitly as
\small{\begin{equation} \label{brcombrantPbf}
\begin{array}{cc}
\underline{\Delta}(B_{i}^{\pm}) = 1 \otimes B_{i}^{\pm} + B_{i}^{\pm} \otimes 1  & \underline{S}(B_{i}^{\pm}) = - B_{i}^{\pm} \\
    \\
\underline{\Delta}(F_{j}^{\pm}) = 1 \otimes F_{j}^{\pm} + F_{j}^{\pm} \otimes 1 &  \underline{S}(F_{j}^{\pm}) = - F_{j}^{\pm} \\
    \\
\underline{\Delta}(\{ B_{k}^{\epsilon}, B_{l}^{\eta} \}) = 1 \otimes \{ B_{k}^{\epsilon}, B_{l}^{\eta} \} + \{ B_{k}^{\epsilon}, B_{l}^{\eta} \} \otimes 1 &
\underline{S}(\{ B_{k}^{\epsilon}, B_{l}^{\eta} \}) = - \{ B_{k}^{\epsilon}, B_{l}^{\eta} \}  \\
   \\
\underline{\Delta}([ F_{\alpha}^{\epsilon}, F_{\beta}^{\eta} ]) = 1 \otimes [ F_{\alpha}^{\epsilon}, F_{\beta}^{\eta} ] + [ F_{\alpha}^{\epsilon}, F_{\beta}^{\eta} ] \otimes 1  &
\underline{S}([ F_{\alpha}^{\epsilon}, F_{\beta}^{\eta} ]) = - [ F_{\alpha}^{\epsilon}, F_{\beta}^{\eta} ] \\
    \\
\underline{\Delta}(\{ F_{\alpha}^{\epsilon}, B_{k}^{\eta} \}) = 1 \otimes \{ F_{\alpha}^{\epsilon}, B_{k}^{\eta} \} + \{ F_{\alpha}^{\epsilon}, B_{k}^{\eta} \} \otimes 1   &
\underline{S}(\{ F_{\alpha}^{\epsilon}, B_{k}^{\eta} \}) = - \{ F_{\alpha}^{\epsilon}, B_{k}^{\eta} \}
\end{array}
\end{equation}}
for all $i, j, k, l, \alpha, \beta = 1,2,... \ $ and for all $\epsilon, \eta = \pm$. We also have $\underline{\varepsilon}(x) = 0 $ $\forall$ $x \in P_{BF}$.

\paragraph{\textbf{On subalgebras of $P_{BF}$}}

The linear subspace of the Relative Parabose set $P_{BF}$ (or of the ($\mathbb{Z}_{2} \times \mathbb{Z}_{2}, \theta$)-Lie algebra $L_{\mathbb{Z}_{2} \times \mathbb{Z}_{2}}$)
spanned by the elements of the form $\{ B_{k}^{\epsilon}, B_{l}^{\eta} \}$, $[ F_{\alpha}^{\epsilon}, F_{\beta}^{\eta} ]$ and $\{ F_{\alpha}^{\epsilon}, B_{k}^{\eta} \}$
for all $k, l, \alpha, \beta = 1,2,... \ $ and for all $\epsilon, \eta = \pm$ is a $\mathbb{Z}_{2}$-graded Lie algebra (or equivalently a Lie superalgebra). The UEA of this Lie superalgebra is a subalgebra of $P_{BF}$.

Let us see this last statement in a little more detail: According to the first statement of this section, the elements $\{ B_{k}^{\epsilon}, B_{l}^{\eta} \}$, $[ F_{\alpha}^{\epsilon}, F_{\beta}^{\eta} ]$ and $\{ F_{\alpha}^{\epsilon}, B_{k}^{\eta} \}$ $\forall$ $k, l, \alpha, \beta = 1,2,... \ $ and $\forall$ $\epsilon, \eta = \pm$ span the $L_{00} \oplus L_{01}$ subspace of the $L_{\mathbb{Z}_{2} \times \mathbb{Z}_{2}} = L_{00} \oplus L_{01} \oplus L_{10} \oplus L_{11}$.
Now it suffices to notice that the subset $\{(0,0), (0,1)\}$ of the $\mathbb{Z}_{2} \times \mathbb{Z}_{2}$ group is a subgroup isomorphic to the $\mathbb{Z}_{2}$ group as we can see from the multiplication tables of the corresponding groups (written in the additive notation):
$$
\begin{array}{ccc}
\begin{array}{c}
\mathbf{\mathbb{Z}_{2} \times \mathbb{Z}_{2} \cong \mathbb{Z}_{2} \oplus \mathbb{Z}_{2}} \ \textbf{group}:   \\  \\
\begin{array}{|c||c|c|c|c|}
  \hline
       \textbf{+}         & \mathbf{(0,0)}  & \mathbf{(0,1)} & \mathbf{(1,0)}  &  \mathbf{(1,1)} \\ \hline \cline{1-5}
 \mathbf{(0,0)} & (0,0) & (0,1) & (1,0)   &   (1,1)            \\   \hline
 \mathbf{(0,1)} & (0,1) & (0,0) & (1,1)   &  (1,0) \\   \hline
 \mathbf{(1,0)} & (1,0) & (1,1) & (0,0)   &   (0,1)     \\   \hline
 \mathbf{(1,1)} & (1,1) & (1,0) & (0,1)   &   (0,0)      \\
\hline
\end{array}
\end{array}     &  \leftrightsquigarrow   &  \begin{footnotesize}
                                             \begin{array}{c}
                                             \{(0,0), (0,1)\} \ is \ a \ subgroup \\
                                             of \ the \ \mathbb{Z}_{2} \times \mathbb{Z}_{2} \ group \\
                                             isomorphic \ to \ the \\
                                             \mathbf{\mathbb{Z}_{2} = \{0, 1\} \ \textbf{group}}:  \\   \\
                                             \begin{array}{|c||c|c|}
                                               \hline
                                               \textbf{+} & \mathbf{0} & \mathbf{1} \\   \hline \cline{1-3}
                                               \mathbf{0} & 0 & 1 \\   \hline
                                               \mathbf{1} & 1 & 0 \\
                                               \hline
                                             \end{array}
                                             \end{array}
                                             \end{footnotesize}
\end{array}
$$
and that the restriction of the commutation factor \eqref{colfunctRelParSe} on $\{(0,0), (0,1)\}$ coincides (as a function) with the usual commutation factor (on $\mathbb{Z}_{2}$) of Lie superalgebras.

\section{Paraparticle Realizations of Lie superalgebras}

For detailed computations and proofs of the results presented in this section one can see \cite{KaDaHa}.

Let $L=L_{0} \oplus L_{1}$ be any complex Lie superalgebra of either finite or infinite dimension and let $V = V_{0} \oplus V_{1}$ be a finite
dimensional, complex, super-vector space i.e. $dim_{\mathbb{C}}V_{0} = m$ and $dim_{\mathbb{C}}V_{1} = n$. If $V$ is the carrier space for a
super-representation (or: a $\mathbb{Z}_{2}$-graded representation) of $L$, this is equivalent to the existence of an homomorphism
$P: \mathbb{U}(L) \rightarrow \mathcal{E}nd_{gr}(V)$ of assoc. superalgebras, from $\mathbb{U}(L)$ to the algebra $\mathcal{E}nd_{gr}(V)$ of $\mathbb{Z}_{2}$-graded
linear maps on $V$.
For any homogeneous element $z \in L$ the image $P(z)$ will be a  $(m+n)\times (m+n)$ matrix of the form
\begin{equation}
\begin{array}{ccc}
 P(z)= \left(\begin{array}{c|c}
              A(z) & B(z) \\ \hline
              C(z) & D(z)
             \end{array}\right)    &  \begin{array}{c}
                                      \nearrow  \\ \\
                                      \searrow
                                      \end{array}   & \begin{array}{c}
                                                      P(X)=
                                                            \left(\begin{array}{ccc}
                                                            A(X) & 0 \\
                                                            0 & D(X)
                                                            \end{array}\right)
                                                            \ \ (if \ z = X \rightsquigarrow \underline{even})
                                                            \\  \\  \\
                                                      P(Y) =
                                                            \left(\begin{array}{ccc}
                                                            0 & B(Y) \\
                                                            C(Y) & 0
                                                            \end{array}\right)
                                                           \ \ (if \ z = Y \rightsquigarrow \underline{odd})
                                                      \end{array}
\end{array}
\end{equation}
where the complex submatrices $A_{m\times m}$, $B_{m\times n}$, $C_{n\times m}$, $D_{n\times n}$,  of $P_{(m+n)\times (m+n)}$ constitute the partitioning ($\mathbb{Z}_{2}$-grading) of the representation.

The linear map $J_{P_{BF}}: L \rightarrow P_{BF}$ defined by
\begin{equation} \label{JordgenPBFeqeven}
X_{i} \mapsto J_{P_{BF}}(X_{i}) = \frac{1}{2} \sum_{k,l=1}^{m}A_{kl}(X_{i}) \{ B_{k}^{+}, B_{l}^{-} \} + \frac{1}{2} \sum_{\alpha,\beta=1}^{n}D_{\alpha\beta}(X_{i}) [ F_{\alpha}^{+}, F_{\beta}^{-} ] \\
\end{equation}
for any even element ($Z=X_{i}$) of an homogeneous basis of $L$ and by
\begin{equation} \label{JordgenPBFeqodd}
Y_{j} \mapsto J_{P_{BF}}(Y_{j}) = \frac{1}{2} \sum_{k=1}^{m}\sum_{\alpha=1}^{n} \Big( B_{k,\alpha}(Y_{j}) \{ B_{k}^{+}, F_{\alpha}^{-} \}
+ C_{\alpha,k}(Y_{j}) \{ F_{\alpha}^{+}, B_{k}^{-} \} \Big) \\
\end{equation}
for any odd element ($Z=Y_{j}$) of an homogeneous basis of $L$, can be extended to an homomorphism of associative algebras
\begin{equation}
J_{P_{BF}}: \mathbb{U}(L) \rightarrow \mathbb{U}(L_{00} \oplus L_{01}) \subset P_{BF}
\end{equation}
between the universal enveloping algebra $\mathbb{U}(L)$ of the Lie superalgebra $L$
and the relative parabose set $P_{BF}$, in other words it constitutes a \textbf{realization of $L$ with paraparticles}. The statement is fully justified by the
fact that the linear map $J_{P_{BF}}$ preserves (see \cite{KaDaHa}) for all values $i, j, p, q = 1,2,...$ the Lie superalgebra brackets:
\begin{equation}
\begin{array}{c}
J_{P_{BF}}(\big[ X_{i}, X_{j} \big]) = \big[ J_{P_{BF}}(X_{i}), J_{P_{BF}}(X_{j}) \big]   \\ \\
J_{P_{BF}}(\big[ X_{i}, Y_{p} \big]) = \big[ J_{P_{BF}}(X_{i}), J_{P_{BF}}(Y_{p}) \big]  \\ \\
J_{P_{BF}}(\{ Y_{p}, Y_{q} \}) = \{ J_{P_{BF}}(Y_{p}), J_{P_{BF}}(Y_{q}) \}
\end{array}
\end{equation}
Furthermore, the above constructed homomorphism of associative algebras $J_{P_{BF}}: \mathbb{U}(L) \rightarrow P_{BF}$, is an homomorphism
of super-Hopf algebras (equivalently an homomorphism of $\mathbb{Z}_{2}$-graded Hopf algebras) between $\mathbb{U}(L)$ and the $\mathbb{U}(L_{00} \oplus L_{01})$ $\mathbb{Z}_{2}$-graded subalgebra of $P_{BF}$. To see this it is enough to compute (see \cite{KaDaHa}) the rhs and the lhs of the following
\begin{equation}
\begin{array}{ccccc}
\underline{\Delta} \circ J_{P_{BF}} = (J_{P_{BF}} \otimes J_{P_{BF}}) \circ \Delta_{L}  &  & \underline{\varepsilon} \circ J_{P_{BF}} = \varepsilon_{L}   &    &
\underline{S} \circ J_{P_{BF}} = J_{P_{BF}} \circ S_{L}
\end{array}
\end{equation}
where $\Delta_{L}, \varepsilon_{L}, S_{L}$ are the Lie superalgebra super Hopf structure maps and $\underline{\Delta}, \underline{\varepsilon}, \underline{S}$ are the braided group structure maps \eqref{brcombrantPbf} of the Relative Parabose Set $P_{BF}$.

\section{Concluding Remarks - Future Prospects}

The Lie superalgebra paraparticle realizations presented in this article constitute direct generalizations of the Jordan-Schwinger map and of realizations presented in works such as \cite{DaKa, Hong, Kad} as well. Our results, generalize these previous works in various aspects: We use Lie superalgebras (instead of Lie algebras), the target algebra consists of the mixed paraparticle system $P_{BF}$ (instead of using simply bosons or fermions or some supersymmetric mixture of them), our formal expressions in \eqref{JordgenPBFeqeven}, \eqref{JordgenPBFeqodd} make use of paraparticle number operators (instead of particle number operators) and finally our results are applicable for both finite and infinite dimensional Lie superalgebras as well.

Similarly to the J.S. map which initiated developments in the representation theory of Lie algebras (construction of the symmetric and the antisymmetric representations of LA), the paraparticle realizations presented in this article can be utilized for pure mathematical purposes as well: Given representations of $P_{BF}$, \eqref{JordgenPBFeqeven}, \eqref{JordgenPBFeqodd} can be the starting point for the construction of possibly new representations of an arbitrary Lie superalgebra, a topic which -at least in the infinite dimensional case- is virtually unexplored. Of course this presupposes the construction of at least the Fock-like representations of $P_{BF}$ which constitutes a hard task on its own.

Finally let us close this work, by outlining a virtual application of the paraparticle realizations \eqref{JordgenPBFeqeven}, \eqref{JordgenPBFeqodd}, which -partially- provided us with the motivation for developing them. It has to do with attempts of the first and the third author of this paper to determine a specific class of solutions of the Skyrme model \cite{Sky}: The Euler-Lagrange equations of the Skyrme model read:
($\vec{\mathbf{L}}_{\mu}$: Maurer-Cartan covariant vectors)
$$
\partial_{\mu} \big( F_{\pi}^{2}\vec{\mathbf{L}}_{\mu} + \dfrac{1}{e} \big[ \vec{\mathbf{L}}_{\nu}, [\vec{\mathbf{L}}_{\mu}, \vec{\mathbf{L}}_{\nu}] \big] \big) = 0
$$
If we impose ``parafermionic'' conditions of the form ($f_{\mu\nu\rho}^{\sigma} \in \mathbb{R} \ or \ \mathbb{C}$)
$$
\big[ [\vec{\mathbf{L}}_{\mu}, \vec{\mathbf{L}}_{\nu} \big], \vec{\mathbf{L}}_{\rho}] = f_{\mu\nu\rho}^{\sigma}\vec{\mathbf{L}}_{\sigma}
$$
between the $\vec{\mathbf{L}}_{\mu}$, enforcing them to be generators of a parafermion-like algebra, then realizing the Maurer-Cartan covariant vectors $\vec{\mathbf{L}}_{\mu}$ in terms of some paraparticle algebra, may prove a helpful intermediate step in determining the constants $f_{\mu\nu\rho}^{\sigma}$. The next step will be the computation of the restrictions imposed on the forms of the fields $U(x)$, implied by the above mentioned ``parafermionic'' conditions. Finally, if the produced forms will prove acceptable, we will proceed in computing the representations of the parafermion-like generated algebra and the physics produced on these representations.

The above constitute part of projects we are already working in and hopefully we will report progress in the near future.


\begin{theacknowledgments}
{\small
KK would like to thank the whole staff of \textsc{Ifm}, \textsc{Umsnh} for providing encouragement and a stimulating atmosphere while this research was in progress. He also acknowledges the postdoctoral grant \textsc{Conacyt}/No. J60060 which partially supported his stay in Michoacan. The research of AHA was supported by grants \textsc{Cic} 4.16 and \textsc{Conacyt}/No. J60060; he is also grateful to \textsc{Sni}.
}
\end{theacknowledgments}

}

\end{document}